\documentclass[aps,prd,preprint,preprintnumbers,unsortedaddress,superscriptaddress,showpacs,nofootinbib]{revtex4-1}
\pdfoutput=1
\usepackage{graphicx}
\usepackage{amsmath}
\usepackage{amsfonts}
\usepackage{amssymb}
\usepackage{color}

\usepackage[colorlinks, citecolor=blue,anchorcolor=red,menucolor=red, linkcolor=red,filecolor=red,runcolor=red,urlcolor=blue,frenchlinks=red]{hyperref}

\usepackage{bm}
\usepackage[section]{placeins}
\usepackage{booktabs}

\begin{document}
\arraycolsep1.5pt

\title{Helicity amplitudes in $B \to  D^{*} \bar{\nu} l$  decay}


\author{L.~R.~Dai}
\email{dailr@lnnu.edu.cn}
\affiliation{Department of Physics, Liaoning Normal University, Dalian 116029, China}
\affiliation{Departamento de F\'isica Te\'orica and IFIC, Centro Mixto Universidad de Valencia-CSIC,
Institutos de Investigac\'ion de Paterna, Aptdo. 22085, 46071 Valencia, Spain
}

\author{E. Oset}
\email{oset@ific.uv.es}
\affiliation{Departamento de F\'isica Te\'orica and IFIC, Centro Mixto Universidad de Valencia-CSIC,
Institutos de Investigac\'ion de Paterna, Aptdo. 22085, 46071 Valencia, Spain
}

\date{\today}
\begin{abstract}
We use a recent formalism of the weak hadronic reactions that maps the transition matrix elements at the quark level into hadronic matrix elements, evaluated with an  elaborate angular momentum algebra
 that allows finally to write the weak matrix elements  in terms of easy analytical formulas. In particular they appear explicitly  for the different spin third components of the vector mesons involved.
 We extend the  formalism  to a general case, with the operator $\gamma^\mu -\alpha\gamma^\mu \gamma_5$, that can accommodate different models beyond the standard model and study in detail the
 $B \to  D^{*} \bar{\nu} l$ reaction for the different  helicities of the $D^*$. The results are shown for each amplitude in terms of the $\alpha$ parameter that  is different for each model. We  show that
 $\frac{d \Gamma}{d M_{\rm inv}^{(\nu l)}}$  is very different for the different components $M=\pm 1, 0$ and in particular the magnitude   $\frac{d \Gamma}{d M_{\rm inv}^{(\nu l)}}|_{M=-1} -\frac{d \Gamma}{d M_{\rm inv}^{(\nu l)}}|_{M=+1} $ is very sensitive to the $\alpha$ parameter, which suggest to use this   magnitude to test different models beyond the standard model.  We also compare our results with the standard model and find very similar results, and practically identical at the end point of  $M_{\rm inv}^{(\nu l)}= m_B- m_{D^*}$.
 \end{abstract}

\maketitle

\section{Introduction}

Semileptonic decays of hadrons have been thoroughly studied and have brought much  information  on the nature of weak interactions and some aspects of QCD \cite{browder,isgur,isgur2,Wirbel:1988ft,neubert,gambino,ecker,neubert2,nieves,tran,a1,a2,lv1,lv2,korner,korner2}.
The relative good control  of the reactions within the standard model (SM) has led to new work searching for evidence  of new physics beyond the standard model (BSM) \cite{Antonelli:2009ws,Fajfer,German}.

One of the magnitudes that has captured  attention as a source of information of new physics BSM is the polarization of vector mesons in $B$ decays. One intriguing feature
was observed  in the $B \to \phi K^*$ decays,   where naively it was  expected  that the transverse amplitudes  would be highly suppressed while the experiment showed equal strength  for longitudinal  and transverse polarizations \cite{aubert}.   Theoretical papers have followed \cite{data,data2}, as well as new experimental measurements  on related reactions, like $B^0_s \to \phi\phi$ \cite{altonen}, $B^+ \to \rho^0 K^{*+}$ \cite{cldamo}, $B^0_s \to K^{*0} \bar{K}^{*0} $ \cite{lhcb}, which had been addressed in papers  dealing with $B \to VV$ decays \cite{kagan,beneke}, $B \to V T$ decays \cite{data3}, and  some particular  reactions as the $B_{(s)} \to D^{(*)}_{(s)} \bar{D}_s^*$  \cite{xuwang}.
  More recently the topic has caught  up in studies  of weak decays into a vector and two leptons as the experiments on $B \to K^* l^+ l^-$ \cite{lees},  $B \to K^* l^+ l^-$ \cite{belle},
$B^0 \to K^{*0} \mu^+ \mu^-$ \cite{cdf,cms,aaij}, and theoretical works on  $B \to K^* \nu \bar{\nu}$ \cite{gudrum,buras}, $B \to K^* l^+ l^-$ \cite{buras}, $B \to K^{*0} l^+ l^-$ \cite{lu1},
$B \to K_J^* l^+ l^-$ \cite{lu2}  and $B \to K_2^* \mu^+ \mu^-$ \cite{lu3}.

  In the present work we retake this line of research and study the polarization amplitudes  in semileptonic $\bar{B} \to V  \bar{\nu} l$  decays, applied in particular to the
$\bar{B} \to D^*  \bar{\nu} l$  reaction. We look at the problem from a different perspective  to  the conventional works  where the formalism   is based on a parametrization of the decay amplitudes
in terms of certain structures involving  Wilson coefficients and form factors. A different approach  was followed  recently in the study of $B$ or $D$  weak decays into two pseudoscalar mesons, one  vector and a    pseudoscalar
and two vectors \cite{liang}. Starting from the operators of the standard model at the quark level, a mapping is done to the hadronic level and  the detailed  angular momentum algebra of the different processes is carried out leading to very simple analytical formulas  for the amplitudes. By means of that,  reactions like $\bar{B}^0 \to D^{-}_s D^+, D^{*-}_s D^+, D^{-}_s D^{*+}, D^{*-}_s  D^{*+}$, and others, can be related up to a global
form factor that cancels in ratios by virtue of heavy quark symmetry. The approach proves very successful in the heavy quark sector and,  due to  the  angular momentum  formalism used, the amplitudes
are generated explicitly  for different  third components of the spin of the  vectors involved. In view of this, the formalism is ideally suited to study polarizations in these
type of decays.

Work along the line of \cite{liang} is also done in  \cite{dai} in the study of the semileptonic $B, B^*,D, D^*$ decays into $\bar{\nu} l$   and a
pseudoscalar or vector meson. Once  again, we can relate different  reactions up to a global  form factor.  If one wished to relate the amplitudes of different spin third components for the
same process, the form factor cancels  in the ratio and  the formalism  makes predictions for  the standard model  without any free parameters.

 In the present work we extend the formalism  and  allow a $(\gamma^\mu- \alpha \gamma^\mu \gamma_5)$ structure for the weak hadronic vertex which makes it easy to make predictions for different  values of $\alpha$ that could occur  in different  models  BSM ($\alpha=1$ here for the SM).  We evaluate  different ratios for the $B \to D^* \bar{\nu} l$ reaction.
 Work on this particular reaction, looking  at the helicity amplitudes within the standard model, was done in \cite{nnew1}.
 A recent work on this issue is presented in \cite{alok}  where the  $B \to D^* \bar{\nu}_\tau \tau$  is studied  separating  the longitudinal  and transverse polarizations.
 The same reaction, looking into $\tau$ and $D^*$ polarization, is studied in \cite{nnew2}.
 Helicity amplitudes are also discussed in the related $\bar{B^*} \to P l \bar{\nu}_l$   reactions in the recent paper  \cite{cq}.

 The formalism  of Ref. \cite{dai} produces directly  the amplitudes in terms of  the third component of the $D^*$ spin along
 the $D^*$  direction. This corresponds to helicity   amplitudes of the $D^*$. The formulas are very easy for these amplitudes and allow to  understand analytically the results that one obtains from the
 final computations.  Not only that, but  they indicate which combinations one should take that make the results most sensitive to the parameter $\alpha$ that will differ from  unity for models BSM.

   We find some observables which are  very sensitive to the value of $\alpha$, which should stimulate experimental work  to investigate possible physics BSM.

\section{Formalism}
\begin{figure}[ht]
\includegraphics[scale=0.72]{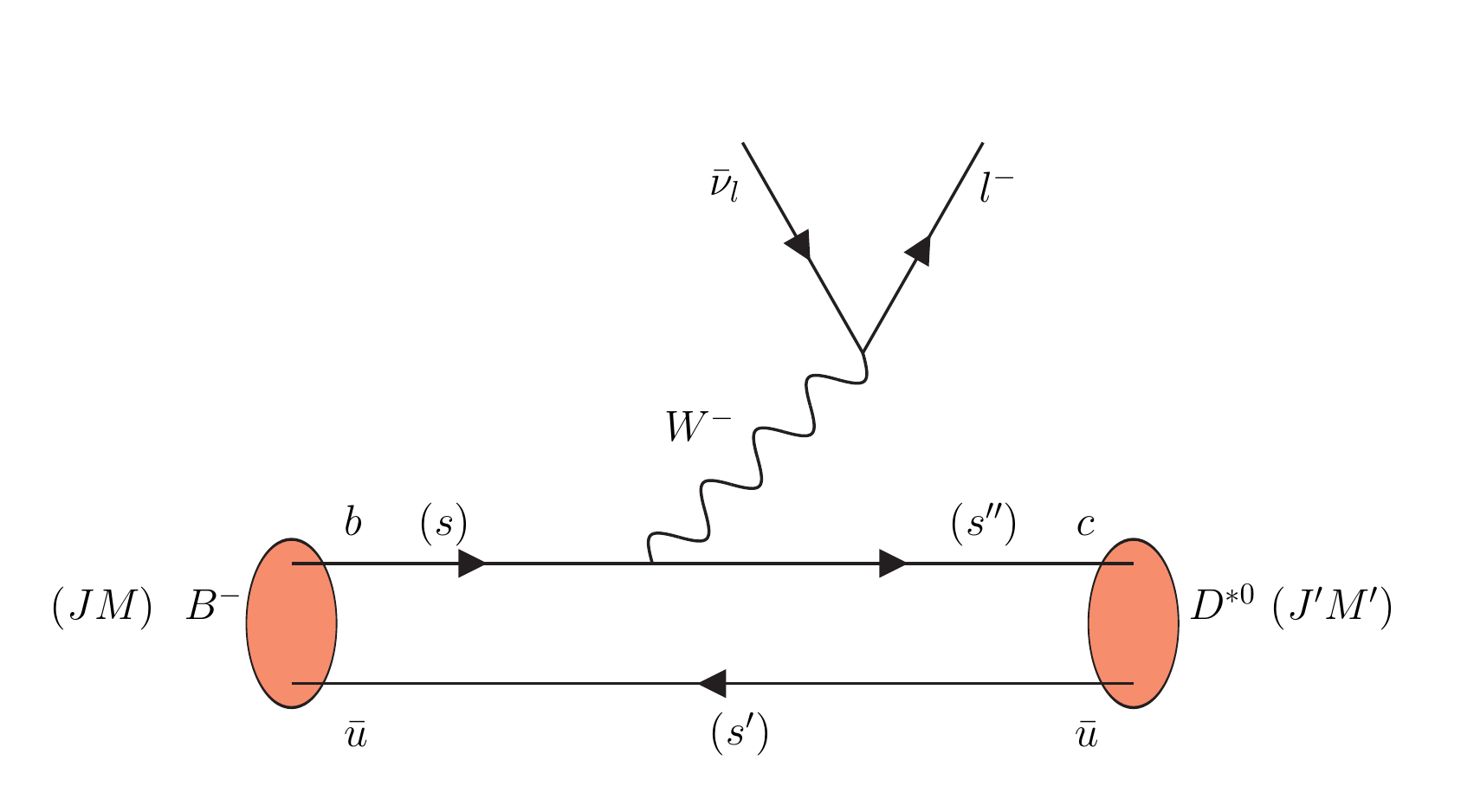}
\caption{Diagrammatic representation of  $B^- \to  D^{*0} \bar{\nu}_{l} l^-$  at the quark level.}
\label{fig:diag}
\end{figure}

We want to study the $B \to  D^{*} \bar{\nu} l$  decay, which  is  depicted  in Fig. \ref{fig:diag}  for  $B^- \to D^{*0} \bar{\nu}_{l} l^-  $

The Hamiltonian  of the weak interaction  is given by
\begin{eqnarray}
H= \mathcal{C} L^\alpha Q_\alpha \,,
\end{eqnarray}
where the $\mathcal{C}$ contains the couplings of the weak  interaction. The constant $\mathcal{C}$  plays no role in our study because we are only concerned about ratios of rates.
 The  leptonic current is given by
\begin{eqnarray}
L^\alpha=\langle {\bar u}_l |\gamma^\alpha(1-\gamma_5)| v_\nu\rangle \,,
\end{eqnarray}
and  the quark current  by
\begin{eqnarray}
Q^\alpha=\langle {\bar u}_c|\gamma^\alpha(1-\gamma_5)|u_b\rangle \,.
\end{eqnarray}

In the evaluation of  $B^- \to  D^{*0} \bar{\nu}_{l} l^-$ decay  we need
\begin{eqnarray}\label{eq:ss}
\overline{\sum} \sum \left|t\right|^2 &=& \sum _{\rm lep~ pol} L^\alpha {L^\beta}^{*} \overline{\sum_{\rm quark}}\sum_{\rm pol}  Q_\alpha Q_\beta^{*} \\ \nonumber \,
&\equiv &{L}^{\alpha\beta} \overline\sum\sum Q_\alpha Q_\beta^{*} \,,
\end{eqnarray}
where $t$ is the transition amplitude, and for simplicity
\begin{eqnarray}
{L}^{\alpha\beta}={\sum _{\rm lep~ pol}} L^\alpha {L^\beta}^{*},
\end{eqnarray}
which can be easily obtained with the result \cite{Navarra}
\begin{eqnarray}\label{eq:ab}
{L}^{\alpha\beta} = 2~ \frac{p_\nu^\alpha p_l^\beta+p_l^\alpha p_\nu^\beta-p_\nu \cdot p_l g^{\alpha\beta}-i \epsilon^{\rho\alpha\sigma\beta} p_{\nu\rho} p_{l\sigma}}{m_\nu m_l} \,,
\end{eqnarray}
where we adopt the Mandl and Shaw normalization  for fermions \cite{mandl}. In Ref. \cite{dai} a study of the meson decays $J M  \to  \bar{\nu}_{l} l J' M'$  was done, where $ J M (J' M')$
 are the modulus and third component of the initial (final) meson spin, and the rates for the different $J, J'$ cases were evaluated. The sum over the $M$ and $M'$ components of Eq. \eqref{eq:ss}
 was done in Ref. \cite{dai}.  Here  we shall keep track of the individual  $M$ and $M'$ contributions.

In evaluating the quark current, we use the ordinary spinors \cite{Itzykson}
\begin{equation}\label{eq:wfn}
u_r=
\widetilde{A}\left(
\begin{array}{c}
\chi_r\\ \widetilde{B} {\bm{\sigma}}\cdot {\bm{p}} \, \chi_r
\end{array}
\right)\,;
\widetilde{A}= \sqrt{\frac{E_p +m}{2\, m}}\,;  \widetilde{B}= \frac{1}{E_p +m}\,,
\end{equation}
where $\chi_r$ are the Pauli bispinors and  $m, p$ and $E_p$ are the  mass, momentum and energy of the quark.
As in Ref. \cite{Navarra} we take
\begin{eqnarray}
\frac{p_b}{m_b}=\frac{p_B}{m_B}\,; \qquad \frac{E_b}{m_b}=\frac{E_B}{m_B}\, ,
\label{eq:ok}
\end{eqnarray}
where $m_B$, $p_B$, and $E_B$ are the mass, momentum and energy  of the $B$ meson,
and the same for the $c$ quark related to the $D^*$ meson.  Theses ratios are tied to the velocity  of the quarks or $B$ mesons and neglect the internal motion of the quarks inside the meson. We evaluate
the matrix elements in the frame where  the   ${\bar\nu}  l$  system  is at rest, where  ${\bm{p}}_B={\bm{p}}_{D^*}={\bm{p}}$,  with $p$ given by
\begin{eqnarray}
p = \frac{\lambda ^{1/2} (m_{B}^2, M_{\rm inv}^{2(\nu l)},  m^2_{D^*})}{2 M_{\rm inv}^{(\nu l)}} \, ,
\label{eq:p}
\end{eqnarray}
where $M_{\rm inv}^{(\nu l)}$ is the invariant mass of the $\nu l$ pair.  By using Eq. \eqref{eq:ok}  we can write
\begin{eqnarray}\label{eq:wfn2}
u_r=
A\left(
\begin{array}{c}
\chi_r\\ B {\bm{\sigma}}\cdot {\bm{p}}_B\, \chi_r
\end{array}
\right)\,; A= \sqrt{\frac{E_B+m_B}{2 m_B}}\,;  B= \frac{1}{m_B+E_B} \, ~~
\end{eqnarray}
and $A',B'$ would be defined for the $D^*$ meson, simply changing the mass in Eq. \eqref{eq:wfn}.

In the present work, we are only interested in the $B^- \to  D^{*0} \bar{\nu}_{l} l^-$  decay, which means $J=0,J'=1$ decay.

As in \cite{dai}  we need to  evaluate ${L}^{\alpha\beta} Q_\alpha Q_\beta^{*}$  which sums  over the polarizations of
$\bar{\nu}_{l} l$, but keeping $M'$ fixed. We have
 \begin{eqnarray} \label{eq:tt}
\sum \left|t\right|^2= L^{00} M_0 M^{*}_0 +  L^{ij}  N_i N_j^{*}  \,.
\end{eqnarray}
where  $M_0$
 \begin{eqnarray}\label{eq:M0}
M_0=-A A^{\prime} (B+B^{\prime}) \, p \, \delta_{M 0} \,\delta_{M^{\prime} 0}
\end{eqnarray}
and $N_i$, written in spherical coordinates,  is
 \begin{eqnarray}\label{eq:Ni}
N_\mu=A A^{\prime} \left \{1+BB^{\prime} p^2 \,(-1)^{-M^{\prime}} +   \sqrt{2}  \right.   \nonumber \, \\
 \left. \times (Bp+B^{\prime} p (-1)^{-M^{\prime}} )\,{\cal C}(1 1 1; M^{\prime},0,M^{\prime})
 \right\}  \delta_{\mu,M^{\prime}} \,\delta_{M 0} \,,~~~~
\end{eqnarray}
with ${\cal C}(\cdots)$ a Clebsch-Gordan coefficient.

In addition to the $p$ dependence  (and hence $M_{\rm inv}^{(\nu l)}$)  of these amplitudes, in \cite{dai} there is an extra form factor coming from
the matrix element of radial $B$ and $D^*$ quark wave functions.
However,  in our approach  we normalize  the different helicity contributions to the
total and the effect of this extra form factor disappears.

The magnitude $L^{ij}  N_i N_j^{*}$ can be written in spherical coordinates  as
\begin{equation}
\sum_{i,j}  L^{ij}  N_i N_j^{*} = \sum_{\alpha,\beta}  (-1)^{\alpha} L^{\alpha \beta} N_{-\alpha} N^{*}_{\beta} \, ,
\end{equation}
and then, following the steps of the appendix of Ref. \cite{dai} we obtain
\begin{itemize}
\item[1)] $M'=0$
\begin{eqnarray}\label{eq:tM0}
 \sum |t|^2  &=&\frac{m^2_l}{m_{\nu} m_l}  \frac{M_{\rm inv}^{2(\nu l)}-m^2_l}{M_{\rm inv}^{2(\nu l)}} \big\{A A' (B+B')p \big\}^2 \\ \nonumber \,
&+&  \frac{2}{m_{\nu} m_l}  \,\left(\widetilde{E}_\nu \widetilde{E}_l +\frac{1}{3} \widetilde{p}_\nu^2 \right) \big\{A A' (1+B B' p^2) \big\}^2    \, .
 \end{eqnarray}
 \item[2)] $M'=1$
\begin{eqnarray}\label{eq:tM1}
\sum |t|^2 &=& \frac{2}{m_{\nu} m_l}  \,\left(\widetilde{E}_\nu \widetilde{E}_l +\frac{1}{3} \widetilde{p}_\nu^2 \right)  \\ \nonumber \,
&\times &  \big\{A A' [(1-B B'p^2) +(B p-B' p)]\big\}^2    \, .
 \end{eqnarray}
 \item[3)] $M'=-1$
\begin{eqnarray}\label{eq:tMm1}
 \sum |t|^2 &=& \frac{2}{m_{\nu} m_l}  \,\left(\widetilde{E}_\nu \widetilde{E}_l +\frac{1}{3} \widetilde{p}_\nu^2 \right) \\ \nonumber \,
&\times & \big\{A A' [(1-B B'p^2) -(B p-B' p)]\big\}^2    \, .
 \end{eqnarray}
 \end{itemize}
where  $\widetilde{p}_\nu$, $\widetilde{E}_\nu$, $\widetilde{E}_l$ are the momentum of the $\bar{\nu}$, its energy and the lepton energy in the rest frame of the $\bar{\nu} l$ system
\begin{eqnarray}
\widetilde{p}_\nu=\widetilde{p}_l=\frac{\lambda^{1/2}(M_{\rm inv}^{2(\nu l)}, m^2_\nu,m^2_l)}{2 M_{\rm inv}^{(\nu l)}}  \nonumber \,, \\
\widetilde{E}_\nu=\frac{M_{\rm inv}^{2(\nu l)}+ m^2_\nu -m^2_l}{2\,M_{\rm inv}^{(\nu l)}}\nonumber \,, \\
\widetilde{E}_l=\frac{M_{\rm inv}^{2(\nu l)}+ m^2_l-m^2_\nu}{2\,M_{\rm inv}^{(\nu l)}}  \,. ~~~~
\end{eqnarray}
$M'$ in Eqs.\eqref{eq:tM0},\eqref{eq:tM1}, and \eqref{eq:tMm1} stands for the third component of the $D^*$ spin  in the direction of $D^*$. Hence these are the helicities of the  $D^*$. Note that in the
boost from the $B$ rest frame to the frame  where $B$ and $D^*$ have the same momentum and $\bar{\nu} l$  are at rest, the direction  of  $D^*$ does not  change and the helicities are the same.
We can see that the sum of these expressions for the three helicities gives  the same result as the sum obtained in Ref. \cite{dai} using properties of Clebsch-Gordan  and Racah coefficients.

\section{Results}

The differential width    is given for  $B \to  D^{*} \bar{\nu} l$   by
\begin{eqnarray}\label{eq:fac18}
\frac{d \Gamma}{d M_{\rm inv}^{(\nu l)}}= \frac{2m_{\nu} 2m_l}{(2\pi)^3} \,\frac{1}{4M^2_B} p'_{D^*} \widetilde{p}_\nu \sum |t|^2  \, ,
\end{eqnarray}
where $p'_{D^*}$ is the $D^*$ momentum in the $B$ rest frame and $\widetilde{p}_\nu$ the  $\bar\nu$ momentum in the $\nu l$ rest frame,
\begin{eqnarray}
p'_{D^*}=\frac{\lambda^{1/2}(m^2_B,M_{\rm inv}^{2(\nu l)}, m^2_{D^*})}{2 m_B}  \, .
\end{eqnarray}
The factor $m_{\nu} m_l$ in the numerator of Eq. \eqref{eq:fac18} is due to the  normalization used in \cite{mandl} and cancels exactly the same factor appearing in the denominator of
Eqs. \eqref{eq:tM0}, \eqref{eq:tM1} and \eqref{eq:tMm1}.

\begin{figure}[ht]
\includegraphics[scale=0.72]{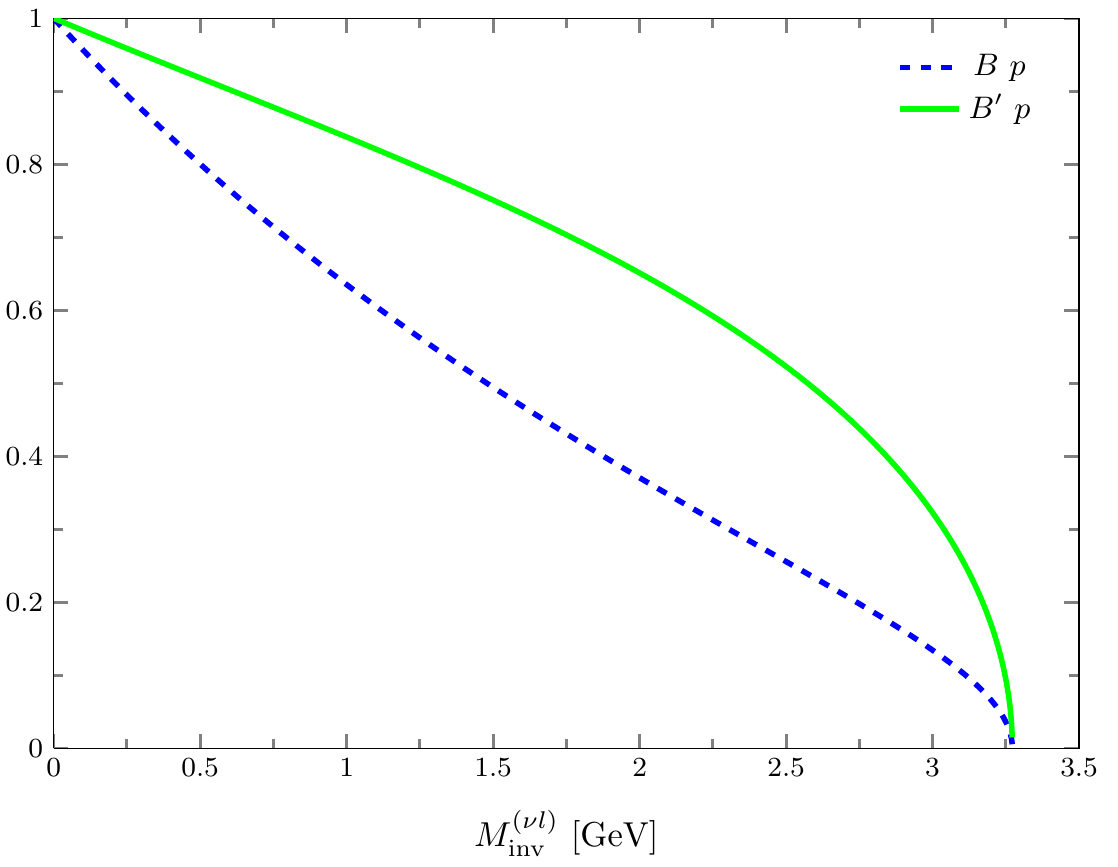}
\caption{Values of $B p$ and $B' p$ as a function of $M_{\rm inv}^{(\nu l)}$ of Eqs. \eqref{eq:p},\eqref{eq:wfn}.}
\label{fig:ab}
\end{figure}

It is interestig to look individually at the distribution of the three third components of the $D^*$ spin. For this we plot $B p$, $B' p$ as a function of $M_{\rm inv}^{(\nu l)}$ in Fig. \ref{fig:ab}. We can see that  $B'p$
is  always bigger than $B p$ and that both $B p$ and $B' p$ go to unity as $M_{\rm inv}^{(\nu l)} \to 0$ (for $m_l=m_\nu=0$).  Moreover,  when $M_{\rm inv}^{(\nu l)}$ goes to its maximum,
then $p \to 0$ and $B p$, $B' p$ go to zero.

Taking into account the behaviour of $B p$ and  $B' p$  depicted in Fig. \ref{fig:ab},  we can see that when $M_{\rm inv}^{(\nu l)} \to 0$ then
$\sum |t|^2 $ goes to $\frac{2 (A A')^2}{m_{\nu} m_l}  \,\left(\widetilde{E}_\nu \widetilde{E}_l +\frac{1}{3} \widetilde{p}_\nu^2 \right) F$, with
$F \to 4, 0, 0$ for $M'=0,1,-1$ respectively, with $(A A')^2\left(\widetilde{E}_\nu \widetilde{E}_l +\frac{1}{3} \widetilde{p}_\nu^2 \right)$ going to a constant.
Conversely,  when $M_{\rm inv}^{(\nu l)} $  goes to  its maximum,  $\sum |t|^2 $ goes to the  same value $\frac{2 (A A')^2}{m_{\nu} m_l}  \,\left(\widetilde{E}_\nu \widetilde{E}_l +\frac{1}{3} \widetilde{p}_\nu^2 \right) $ for  $M'=0,1,-1$ cases.

\begin{figure}[ht]
\includegraphics[scale=0.72]{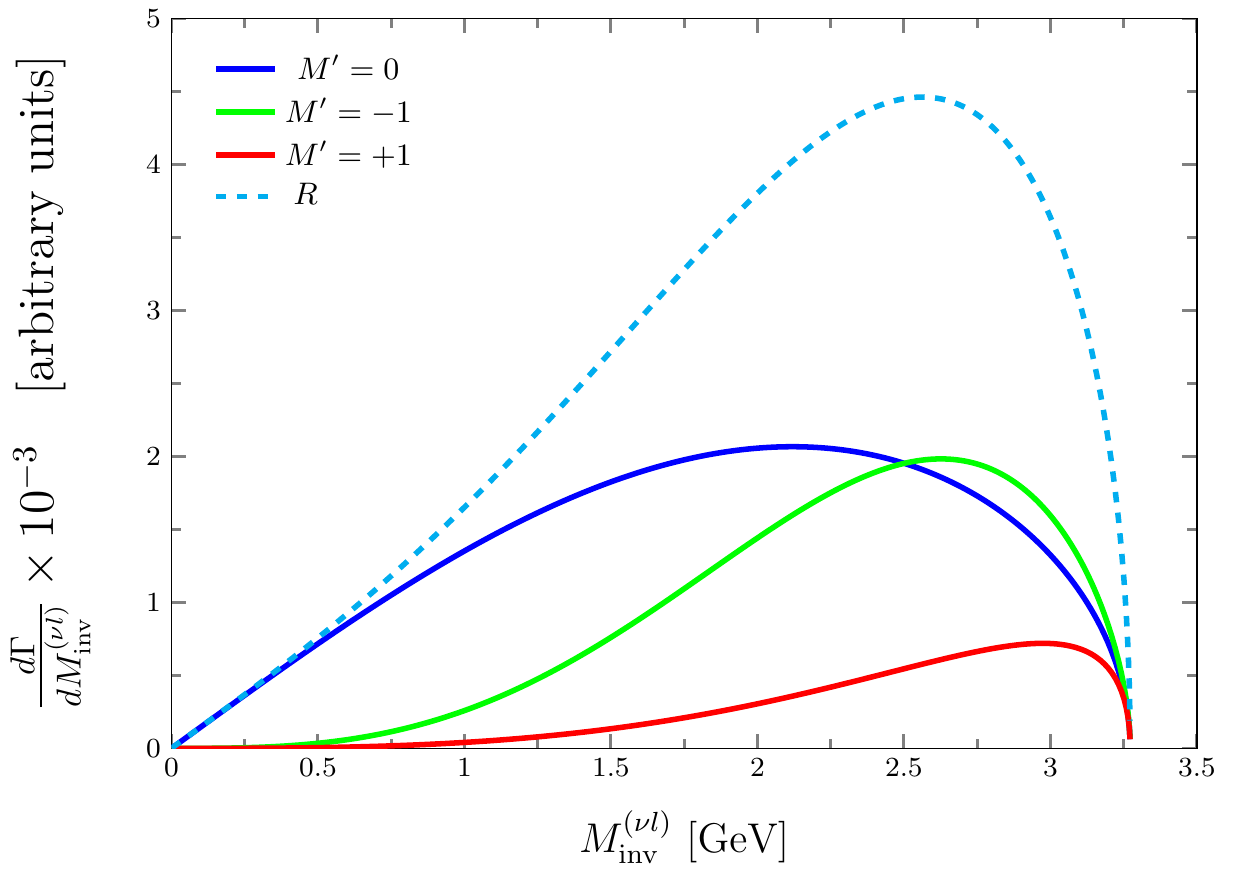}
\caption{ Total  differential width $R$, and individual contributions of $\frac{d \Gamma}{d M_{\rm inv}^{(\nu l)}}|_{M'=0}$,
$\frac{d \Gamma}{d M_{\rm inv}^{(\nu l)}}|_{M'=-1}$, and $\frac{d \Gamma}{d M_{\rm inv}^{(\nu l)}}|_{M'=+1}$.}
\label{fig:dg}
\end{figure}
\begin{figure}[ht]
\includegraphics[scale=0.72]{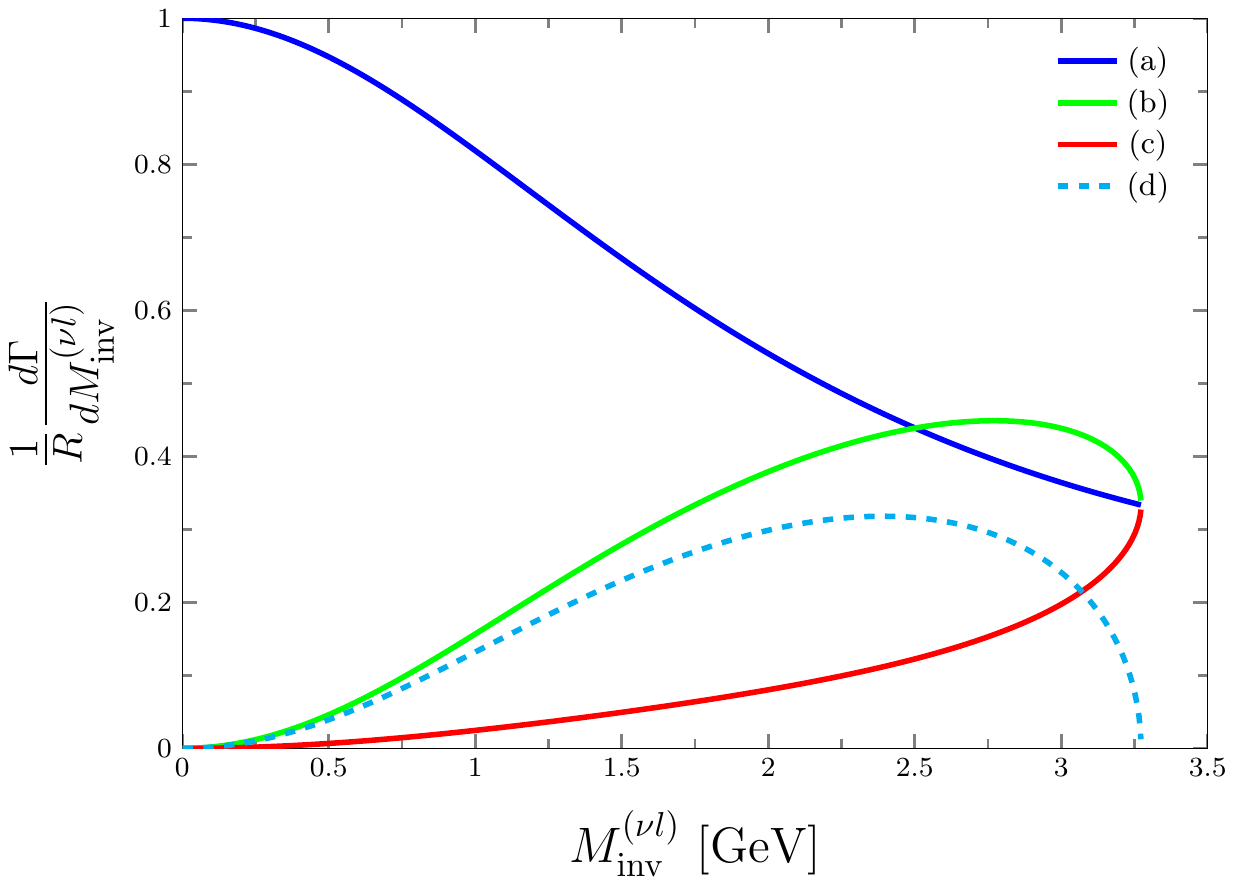}
\caption{The same as Fig. \ref{fig:dg} but for different  ratios. The  lines (a), (b) and (c) show  $\frac{d \Gamma}{d M_{\rm inv}^{(\nu l)}}|_{M'=0}$, $\frac{d \Gamma}{d M_{\rm inv}^{(\nu l)}}|_{M'=-1}$, and
$\frac{d \Gamma}{d M_{\rm inv}^{(\nu l)}}|_{M'=+1}$ respectively, and line (d) denotes the
 difference of  $\frac{d \Gamma}{d M_{\rm inv}^{(\nu l)}}|_{M'=-1}-\frac{d \Gamma}{d M_{\rm inv}^{(\nu l)}}|_{M'=+1}$, all  divided by the total differential width $R$ of Eq. \eqref{eq:R}. }
\label{fig:ro}
\end{figure}

It is  also interesting to see that
\begin{eqnarray}
\overline{\sum}\sum (|t|^2_{M'=-1} -|t|^2_{M'=+1}) = \frac{8}{m_{\nu} m_l}  \,\left(\widetilde{E}_\nu \widetilde{E}_l +\frac{1}{3} \widetilde{p}_\nu^2 \right) \nonumber \, \\
 \times (A A')^2 (1 - B B' p^2)(B' p-B p) \, . ~~~~~~~
\end{eqnarray}
This means that the differential width $d \Gamma/d M_{\rm inv}^{(\nu l)}$ for this difference goes as $(1 - B B' p^2)(B' p-B p) $ and the difference of these two distributions goes to zero,
both  as  $M_{\rm inv}^{(\nu l)} \to 0$ or  $M_{\rm inv}^{(\nu l)} $ going to its  maximum.
We show also these results in Figs.  \ref{fig:dg} and \ref{fig:ro}. The total  differential width is given by
\begin{eqnarray}\label{eq:R}
 R=\frac{d \Gamma}{d M_{\rm inv}^{(\nu l)}}|_{M'=0} + \frac{d \Gamma}{d M_{\rm inv}^{(\nu l)}}|_{M'=-1} +\frac{d \Gamma}{d M_{\rm inv}^{(\nu l)}}|_{M'=+1} \, . ~~~~~
\end{eqnarray}

In Fig. \ref{fig:dg} we  show the individual contribution of each $M'$ and the total. In Fig. \ref{fig:ro} we show the contribution of each $M'$ and the difference
of $M'=-1$  and $M'=+1$, divided by  the total  differential width  $R$.  In this latter figure we can see how fast the  individual $M'=-1$ and $M'=+1$
components go to zero.

In the search for contributions beyond the standard model (SM) one  usually compares some magnitude with experiment and diversions of experiment with respect to the SM predictions
 are seen as a signal of possible new physics. So far the experimental errors do not make  the cases compelling. The present case could offer a good opportunity,  since
the individual contributions for different $M'$ vary appreciably when diverting from the standard model, as we show in the next section.


We also appreciate  in Fig. \ref{fig:ro}  that  the ratio of $ \frac{d \Gamma}{d M_{\rm inv}^{(\nu l)}}|_{M'=-1}-\frac{d \Gamma}{d M_{\rm inv}^{(\nu l)}}|_{M'=+1}$ divided by the
total differential width goes fast to zero for $M_{\rm inv}^{(\nu l)} \to 0$ or  maximum, while individually each of the contributions goes to $\frac{1}{3}$ at the maximum of $M_{\rm inv}^{(\nu l)}$.
In  Fig. \ref{fig:ro} we also see a smooth transition from $1$ to $\frac{1}{3}$  for the $M'=0$ case.    The rapid transition to zero of some of the amplitudes discussed and the
wide change  of values for the (a), (b), (c) and (d) cases in the figure make these magnitudes specially suited to look for extra  contribution beyond  the SM.

To give a further insight into this issue we stress that the reason for the zero strength at $M_{\rm inv}^{(\nu l)} \to 0 $ in the case of $M'=\pm 1$, is tied not to the lepton current, since
we always get  $(A A')^2\left(\widetilde{E}_\nu \widetilde{E}_l +\frac{1}{3} \widetilde{p}_\nu^2 \right)$, which goes to a constant for $M_{\rm inv}^{(\nu l)} \to 0$, but to the quark current.
Indeed, if we look at Eqs. \eqref{eq:M0}, \eqref{eq:Ni}, we can see that both $\frac{M_0}{AA'}$  and $\frac{N_\mu}{AA'} (\mu=0)$  are different from zero for $M'=0$ in the limit of $M_{\rm inv}^{(\nu l)} \to 0$.
However, for $M'=\pm 1$, $M_0=0$ and $N_\mu$  goes to zero in that limit. This said, the models beyond the SM which could provide finite contribution for  $M'=\pm 1$, or a sizeably bigger one,
  are those that go beyond
the $\gamma^\mu- \gamma^\mu\gamma_5$ structure in the quark current, like leptoquarks or right-handed quark currents of the type
$\gamma^\mu+ \gamma^\mu\gamma_5$  \cite{arXiv1511,arXiv1608,gang}.  We discuss this case below.

\section{Consideration of right-handed quark currents}

\begin{figure}[ht]
\includegraphics[scale=0.72]{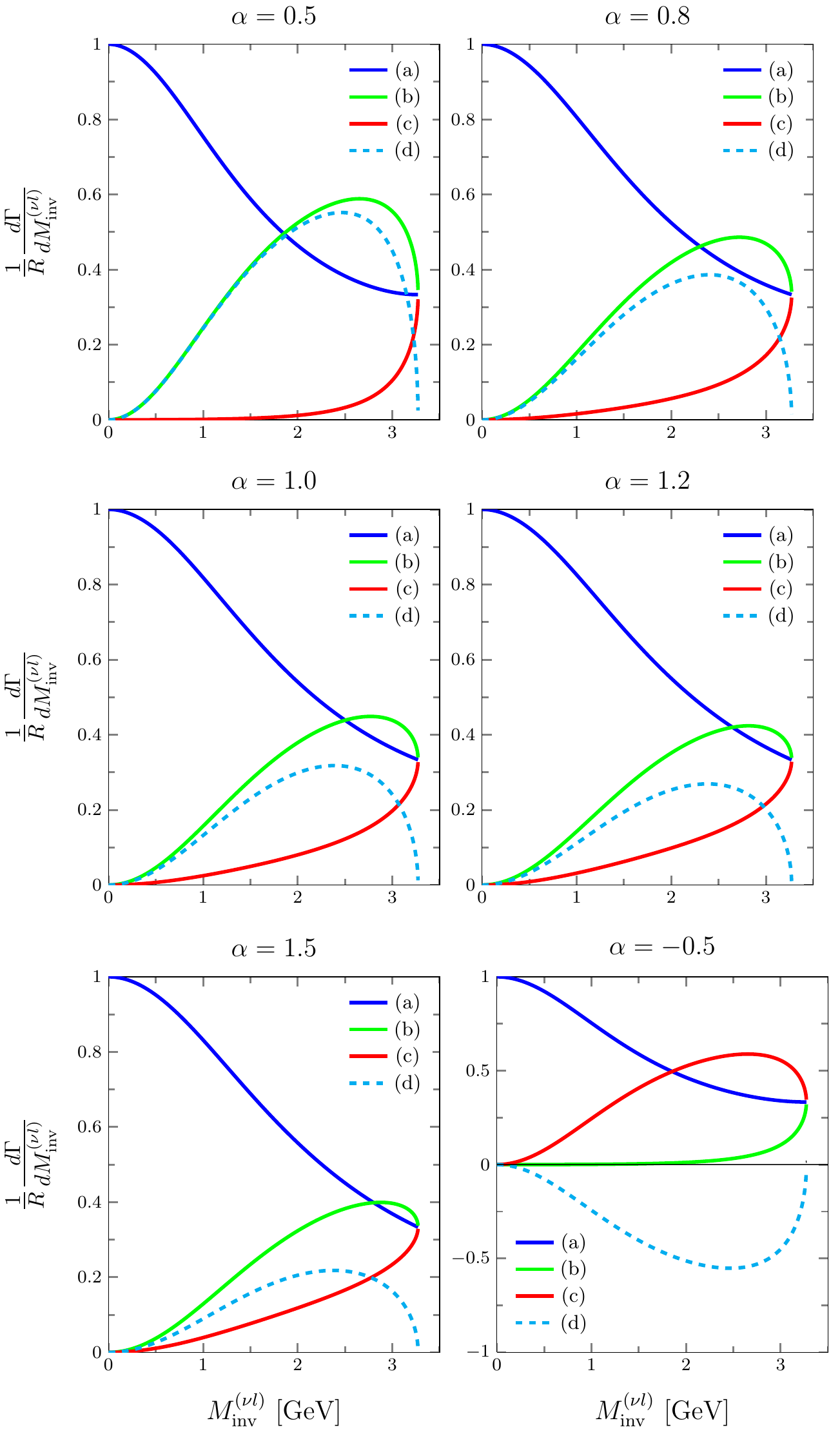}
\caption{The same as Fig. \ref{fig:ro} but for different $\alpha$.  }
\label{nfig:ro}
\end{figure}

The literature about models BSM is large and this is not the place to discuss it. Yet, we would like to mention more recent papers on models which could be easily tested within the present approach,
a minimal gauge extensions of the SM \cite{avelino,virto}, leptoquarks \cite{okada}, scalar leptoquarks \cite{darice,new1},  vector leptoquarks \cite{isidori,luzio,calibbi}, Pati-Salam gauge models \cite{bordone,Fornal,new2}
and right-handed models \cite{German,hedos}.

Some models BSM have quark currents that contain the combination $\gamma^\mu+ \gamma^\mu \gamma_5$.
The models mentioned above could be accommodated with an operator
\begin{eqnarray}
&a(\gamma^\mu -\gamma^\mu \gamma_5)+b(\gamma^\mu+\gamma^\mu \gamma_5) \nonumber \, \\
=&(a+b)\left\{\gamma^\mu -\frac{a-b}{a+b}\gamma^\mu \gamma_5 \right\}  \,.~~~~
\end{eqnarray}
We shall call $\frac{a-b}{a+b}=\alpha$ and  study the   distributions  for different $M'$ as a function of $\alpha$.  We  have thus the operator
\begin{eqnarray}
\gamma^\mu -\alpha\gamma^\mu \gamma_5  \nonumber \,  .
\end{eqnarray}
Using the same formalism  of \cite{dai} it is easy to see the results as a function of $\alpha$. We obtain the following results:
 \begin{equation}\label{neq:M0}
M_0=- A A^{\prime} (B+B^{\prime}) p \delta_{M 0} \,\delta_{M^{\prime} 0} \,\alpha
\end{equation}
and $N_i$ written in spherical coordinates is
\begin{eqnarray}\label{neq:Ni}
N_\mu=A A^{\prime}\left\{\big[1+BB^{\prime} p^2 \,(-1)^{-M^{\prime}}\big]\alpha + \sqrt{2}  \right.   \nonumber \, \\
  \left.  \times [Bp+B^{\prime} p (-1)^{-M^{\prime}} ] \ {\cal C}(1 1 1; M^{\prime},0,M^{\prime}) \right\}  \delta_{\mu,M^{\prime}} \,\delta_{M 0}  \,.~~~~~
 \end{eqnarray}
Then, the different helicity contributions are given by
\begin{itemize}
\item[1)] $M'=0$
\begin{eqnarray}\label{neq:tM0}
 \sum |t|^2  &=&\frac{m^2_l}{m_{\nu} m_l}  \frac{M_{\rm inv}^{2(\nu l)}-m^2_l}{M_{\rm inv}^{2(\nu l)}} \big\{A A' (B+B')p \big\}^2 \alpha^2 \\ \nonumber \,
&+&  \frac{2}{m_{\nu} m_l}  \,\left(\widetilde{E}_\nu \widetilde{E}_l +\frac{1}{3} \widetilde{p}_\nu^2 \right) \big\{A A' (1+B B' p^2) \big\}^2 \alpha^2   \, .
 \end{eqnarray}
 \item[2)] $M'=1$
\begin{eqnarray}\label{neq:tM1}
\sum |t|^2 &= &\frac{2}{m_{\nu} m_l}  \,\left(\widetilde{E}_\nu \widetilde{E}_l +\frac{1}{3} \widetilde{p}_\nu^2 \right) \\ \nonumber \,
&\times & \big\{A A' [(1-B B'p^2)\alpha +(B p-B' p)]\big\}^2    \, .
 \end{eqnarray}
 \item[3)] $M'=-1$
\begin{eqnarray}\label{neq:tMm1}
 \sum |t|^2 &=& \frac{2}{m_{\nu} m_l}  \,\left(\widetilde{E}_\nu \widetilde{E}_l +\frac{1}{3} \widetilde{p}_\nu^2 \right) \\ \nonumber \,
&\times &  \big\{A A' [(1-B B'p^2)\alpha -(B p-B' p)]\big\}^2    \, .
 \end{eqnarray}\end{itemize}

Since  $1 - B B' p^2$ and $B' p-B p$ go individually  to zero for $M_{\rm inv}^{(\nu l)} \to 0$, then we see that  the  $M_{\rm inv}^{(\nu l)}$ distributions for $M'=\pm 1$ still go to zero
in that limit. Yet, the individual contributions  depend strongly on $\alpha$.

In Fig. \ref{nfig:ro}  we show the results of $M'=\pm 1$ and $M'=0$  for different values of $\alpha$. We can see that for $\alpha=0.5$ the contribution of   $M'=+1$ eventually vanishes.
However, for $\alpha=1.5$ it is much bigger and close to the distribution of $M'=-1$.  For $\alpha=-0.5$ the values for $M'=+1$ and $M'=-1$ are exchanged with respect to $\alpha=0.5$ and the $M'=+1$ contribution is much
bigger than the one of $M'=-1$.  Such cases should be  easy to differentiate experimentally.

\begin{figure}[ht!]
\includegraphics[scale=0.72]{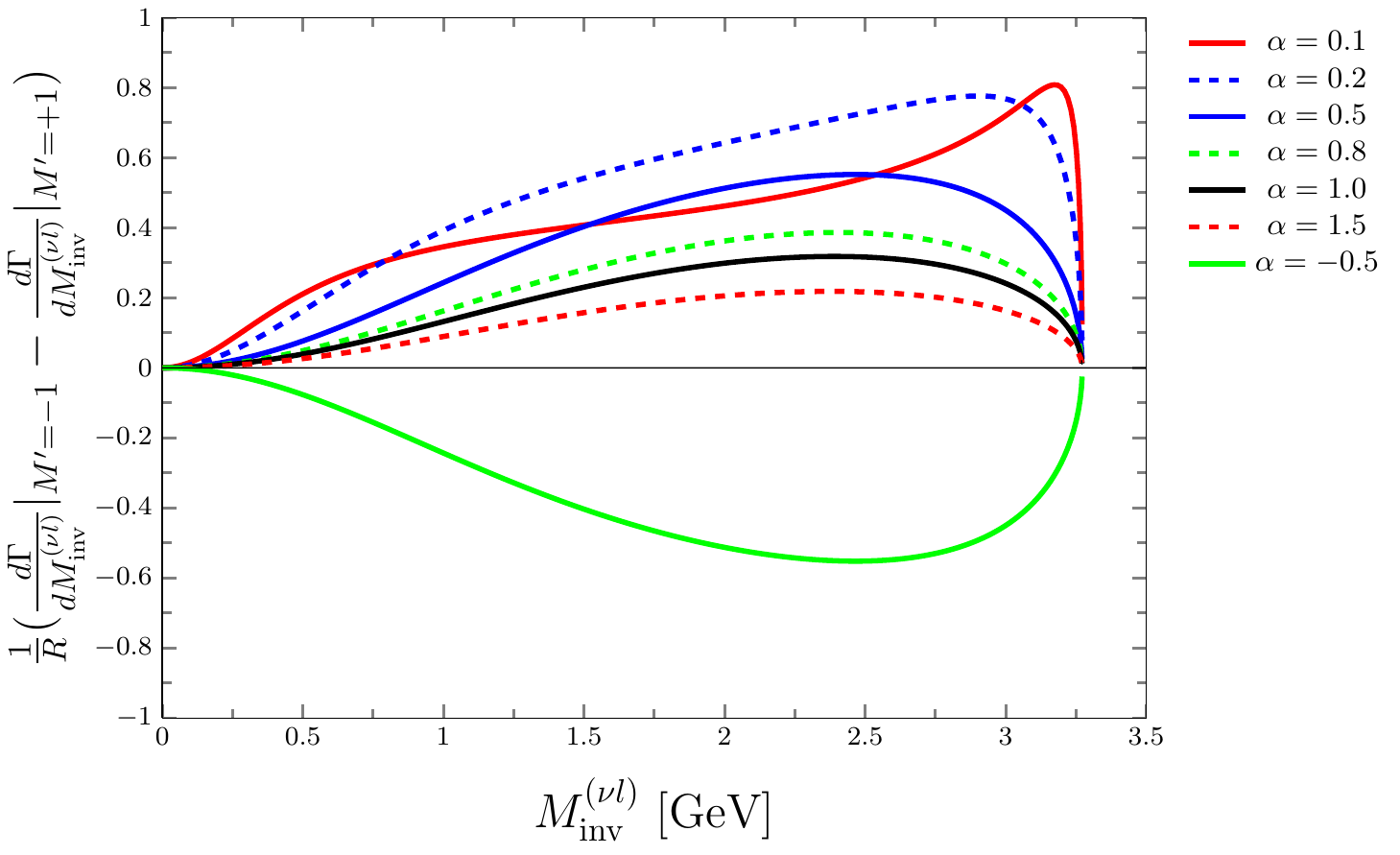}
\caption{ The  difference of  $\frac{d \Gamma}{d M_{\rm inv}^{(\nu l)}}|_{M'=-1}-\frac{d \Gamma}{d M_{\rm inv}^{(\nu l)}}|_{M'=+1}$
of Eq. \eqref{neq:dif} as a function of $\alpha$, divided by the total differential width $R$. }
\label{fig:nro}
\end{figure}

We also see that  now
\begin{eqnarray}\label{neq:dif}
\overline{\sum}\sum (|t|^2_{M'=-1} -|t|^2_{M'=+1}) = \frac{8 \, \alpha}{m_{\nu} m_l}  \,\left(\widetilde{E}_\nu \widetilde{E}_l +\frac{1}{3} \widetilde{p}_\nu^2 \right)     \nonumber \, \\
\times (A A')^2 (1 - B B' p^2)(B' p-B p) \, . ~~~~~~~~
\end{eqnarray}

 Then, it is also interesting  to see what happens for the ratio $\frac{1}{R}(\frac{d \Gamma}{d M_{\rm inv}^{(\nu l)}}|_{M'=-1}-\frac{d \Gamma}{d M_{\rm inv}^{(\nu l)}}|_{M'=+1})$.
 We show these results in  Fig. \ref{fig:nro}.
 We can see that this magnitude keeps rising up as $\alpha$ goes from $1.5$ to  about $0.2$. For $\alpha=0.1$  the shape changes drastically,
 and as $\alpha$ goes to zero it changes very fast. This is because for $\alpha \to -\alpha$ the magnitude changes sign.
 We also show  the value  of the magnitude for $\alpha=-0.5$, which indeed is symmetric of the one for $\alpha=0.5$ with respect to the $M_{\rm inv}^{(\nu l)}$ axis.
It is clear that magnitudes like this, which can change sign from one model to another, should be very useful in the search of contributions BSM.

\section{Connection with the conventional formalism and the standard model}
In Ref. \cite{dai} one relates the weak amplitudes for the ${B} \to D  \bar{\nu} l, D^*  \bar{\nu} l$, ${B}^* \to D  \bar{\nu} l,  D^*  \bar{\nu} l$.  This means that there is only one independent
amplitude for all  these processes.  This is reminiscent of the heavy quark symmetry \cite{falk,nuebertsolo} where all form factors can be cast in terms of only one in the limit of infinite
masses of the mesons. In view of this, let us face this issue here to see the  heavy quark symmetry implicit  in the approach of \cite{dai} which we follow here.

The key point in our approach, which allows us to express the quark matrix elements in terms of the meson variables, is  Eq. \eqref{eq:ok}.  Let is take the first relation
$\frac{p_b}{m_b}=\frac{p_B}{m_B}$. In the  $B$ meson at rest there is a distribution of quark momenta due to the internal motion of the quarks, ${\bm p_{in}}$. If we make
a boost to have  the $B$ with a velocity of $\bm{v}$, we will have

\begin{eqnarray}\label{eq:np}
{p'}_{bL}=\frac{{p}_{in,L}+vp_{in}^0}{\sqrt{1-v^2}}  \,, \qquad p'_{bT}=p_{bT}   \,,    \nonumber \,
\end{eqnarray}
where we have split $\bm{p_{in}}$ into a longitudinal and transverse part along the direction of $\bm{v}$. We can write now
\begin{eqnarray}
\frac{{p}_{bL}}{p_B} &=&\frac{{p}_{in,L}+vp_{in}^0}{\sqrt{1-v^2}} \div \frac{m_B v}{\sqrt{1-v^2}} =\frac{{p}_{in,L}+vp_{in}^0}{m_B v}  \nonumber \, \\
&=&\frac{p_{in}^0}{m_B}+\frac{{p}_{in,L}}{m_B v}=\frac{m_b}{m_B}+\frac{{p}_{in,L}}{m_B v}    \nonumber   \,.
\end{eqnarray}
The relative correction factor is
\begin{eqnarray}
\frac{{p}_{in,L}}{m_B v}\div \frac{m_b}{m_B} =\frac{{p}_{in,L}}{m_b v} \simeq \frac{{p}_{in,L}}{m_B v}    \nonumber \,,
\end{eqnarray}
but since ${p}_{in,L}$ has positive and  negative components the correction is of order
\begin{eqnarray}
\left(\frac{{p}_{in,L}}{m_B v} \right)^2 \simeq \frac{1}{3} \frac{{p}^2_{in}}{m_B^2 v^2}   \,.
\end{eqnarray}
Let us remark that around  $M_{\rm inv}^{(\nu l)}|_{\rm min}$,  $p$ from Eq. \eqref{eq:p} becomes infinite and thus $v=1$. Hence the correction terms are of the order of
$\frac{1}{3} \frac{{p}^2_{in}}{m_B^2 v^2}$. With typical values of $|\bm p_{in}|\simeq 300$ MeV, this is a correction of one permil.  For the $D^*$ meson is a correction of
less than $1\%$. We can make $v$ smaller  as $M_{\rm inv}^{(\nu l)} $ grows and still keep these numbers very small. Certainly, when we go to the end point, for $M_{\rm inv}^{(\nu l)}|_{\rm max}$,
when both $B$ and $D^*$ are at rest,  the argument would fail since $v=0$.  However, in this case  the approximation is equally good since the $Bp$ term is zero and only $A,A'$ matter and
$E_b=m_b$ at the level of $\frac{p^2_{in}}{2m^2_B}$ and $E_B=M_B$, hence $\widetilde{A}$ in Eq. \eqref{eq:wfn} and $A$ in Eq. \eqref{eq:wfn2} are again remarkably close. Incidentally, the transverse components
in the boosted frame lead to a correction of
\begin{eqnarray}
\frac{{p}_{in,T}}{p_B} =\frac{{p}_{in,T} \sqrt{1-v^2}}{m_B v}  \to \frac{2}{3} \frac{{p}^2_{in}(1-v^2)}{m_B v^2} \nonumber \,,
\end{eqnarray}
and their effect is further negligible. One can repeat the argumentation for  the second relation of Eq. \eqref{eq:ok}. This indicates that in  the $\bar{\nu} l$ rest frame, where we evaluate
the matrix elements,  Eq. \eqref{eq:ok} is very accurate. However, it is only exact in the strict limit that $m_B,m_{D^*}$ go to infinite. Hence, it should not be surprising that our method
implements automatically  the symmetries of heavy quark physics.

  In order to test this hypothesis let us  first study the $\bar{B} \to D  \bar{\nu} l$ transition.  We have \cite{isgurwise}
\begin{eqnarray}
\frac{<D, P'|{\cal J}_\mu (0)|B, P>}{\sqrt{m_B m_D}} =( v+v')_\mu h_{+}(w)+(v-v')_\mu h_{-}(w)  \nonumber \,, ~~~
\end{eqnarray}
where $v=\frac{P}{M_B}$, $v'=\frac{P'}{M_D}$ and
\begin{eqnarray}
 w= v v'= \frac{M^2_B+M^2_D-M_{\rm inv}^{2(\nu l)}}{2M_B M_D} \,,
\end{eqnarray}
($M_D \to M_{D^*}$ for the $\bar{B} \to D^*  \bar{\nu} l$  transition).
Similarly (using $\epsilon^{0123}=1$) we have \cite{isgurwise,neubert2}
\begin{eqnarray}
&\frac{<D^*,\lambda,  P'|{\cal J}_\mu (0)|B, P>}{\sqrt{m_B m_{D^*}}} = i \epsilon_{\mu\nu\alpha\beta} (\epsilon^{(\lambda)\nu})^* v^\alpha v^{'\beta} h_V\nonumber \, \\
& -(\epsilon^{(\lambda)*} (w+1) h_{A_1} + (\epsilon^{(\lambda)*} \cdot v) (v_\mu h_{A_2} + v'_\mu h_{A_3})   \nonumber \, .~~~
\end{eqnarray}
In the   heavy quark  limit, with the quark masses going to infinite, one finds \cite{isgurwise,neubert2}
\begin{eqnarray}
&h_+(w)=h_{A_1}(w)=h_{A_3}(w)=h_{V}(w)=\xi (w)   \nonumber \,, \\
&h_{-}(w)=h_{A_2}(w)=0 \,,
\end{eqnarray}
with $\xi (w)$ the Isgur Weise function, and with a certain normalization  of $\cal{J}_\mu$, $\xi (w)$  at the end point, $M_{\rm inv}^{(\nu l)}|_{\rm max}=M_B-M_{D,D^*}  (p=0)$,
\begin{eqnarray}
\xi (w=1) =1 \,.
\end{eqnarray}
This condition appears naturally in the quark model since for $w=1$ the momentum transfer is zero and the wave functions with very large quark  masses  are also equal. Hence the quark transition form
factor is unity.

\begin{figure}[ht!]
\includegraphics[scale=0.72]{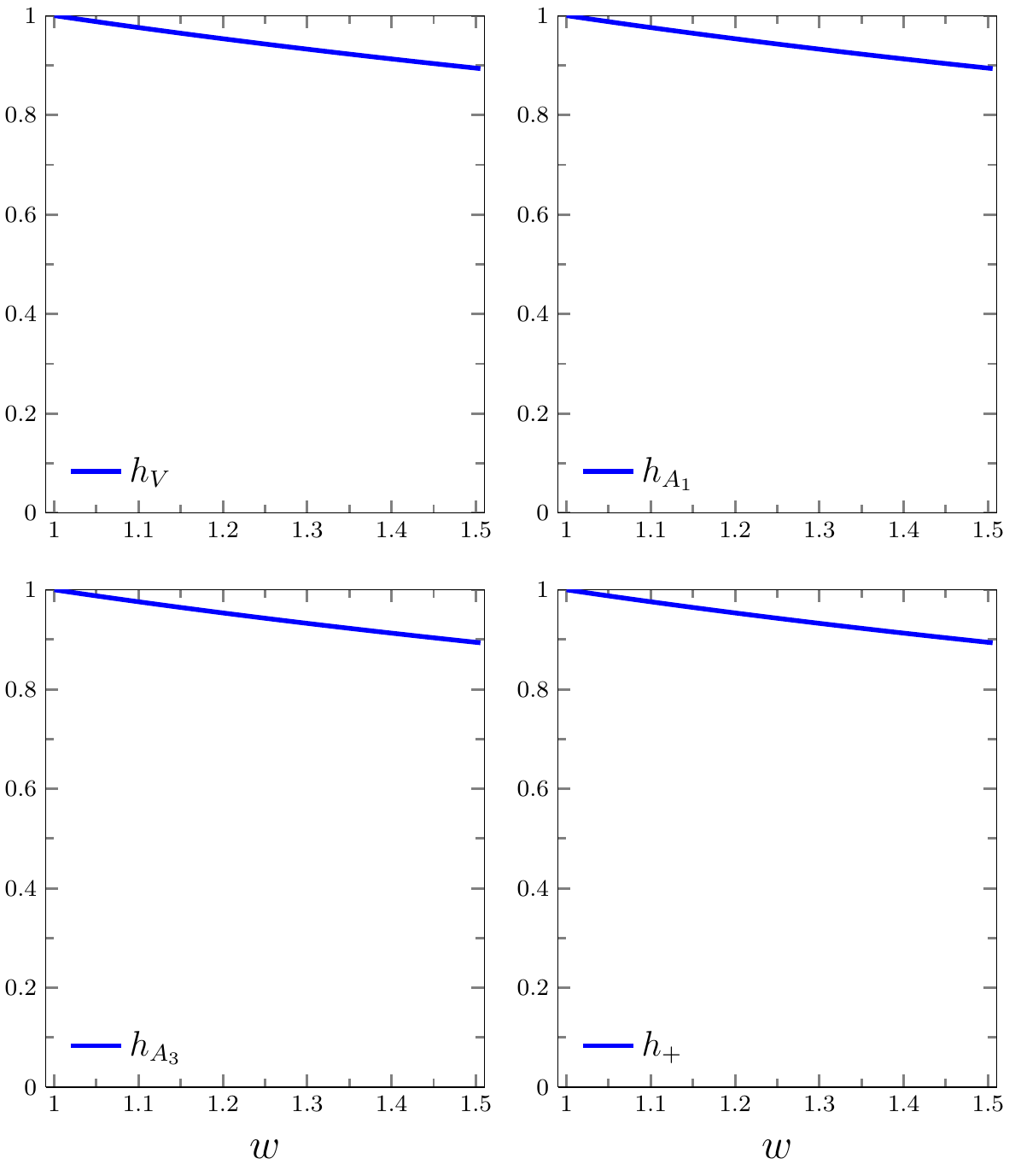}
\caption{$h_{+}$, $h_{A_1}$, $h_{V}$ and $h_{A_3}$ of Eq. \eqref{eq:h} as a function of $w$  normalized to $1$  at $w=1$. }
\label{fig:h}
\end{figure}

We take the $D^*$ polarization  vectors consistent  with our convention in  \cite{dai} for the angular momentum states
\begin{eqnarray}
&M'=0 \,, \qquad  &\epsilon^{(0)\nu}\equiv \left(\frac{p}{M_{D^*}},0,0,\frac{E_{D^*}}{M_{D^*}} \right)    \nonumber \,,\\
&M'=1 \,, \qquad  &\epsilon^{(+)\nu}\equiv   -\frac{1}{\sqrt{2}}\left(0,1,i,0 \right)    \nonumber \,, \\
&M'=-1 \,, \qquad &\epsilon^{(-)\nu}\equiv   \frac{1}{\sqrt{2}}\left(0,1,-i,0 \right)    \,.
\end{eqnarray}
By using these polarization  factors we  compare  the ${\cal J}_0$, ${\cal J}_i$ (${\cal J}_{\tilde{\mu}}$ in spherical basis) matrix elements with $M_0$ and $N_{\tilde{\mu}}$ of the expressions found in \cite{dai}.
\begin{itemize}
  \item [1)] $J=0,J'=0$ ($\bar{B} \to D  \bar{\nu} l$ )
\begin{eqnarray}
 M_0 &=& A A' (1+B B'P^2)  \, \delta_{M 0} \, \delta_{M' 0}  \nonumber \,, \\
 N_{\tilde{\mu}}&=&-A A' (B+B') \, p  \,\delta_{M0}  \,\delta_{M' 0} \, \delta_{\mu 0}  \,.
\end{eqnarray}
\item [2)] $J=0,J'=1$ ($\bar{B} \to D^*  \bar{\nu} l$) \\
$M_0$ and $N_{\tilde{\mu}}$  are given by Eqs. \eqref{eq:M0} and \eqref{eq:Ni}.
\end{itemize}
We find
\begin{eqnarray}\label{eq:h}
h_{+} &=& \frac{\sqrt{m_B m_D}}{m_B+m_D} A A' (B + B')  \nonumber \,, \\
h_{-} &=& 0  \nonumber \,, \\
h_{A_1} &=& \frac{1}{w+1} A A'(1-B B' p^2) \frac{1}{\sqrt{m_B m_{D^*}}}   \nonumber \,, \\
h_{V} &=& \sqrt{m_B m_{D^*}}  A A'\frac{(B-B')}{E_{D^*}-E_B} \nonumber \,, \\
h_{A_2} &=& 0  \nonumber \,, \\
h_{A_3} &=& \frac{M^2_{D^*}M_B}{E_B-E_{D^*}} \frac{A A'}{E_{D^*}\sqrt{m_B m_{D^*}}}  \left\{\frac{1}{M_{D^*}}(1-B B' p^2)-(B+B') \right\} \,.
\end{eqnarray}
Because of our normalization for $\cal{J}_\mu$, all these functions are normalized to the value $\frac{1}{2\sqrt{m_B m_{D^*}}}$ at $w=1~(p=0)$. Multiplying by $2\sqrt{m_B m_{D^*}}$ we find the
form factors  normalized to $1$.  In Fig. \ref{fig:h}  we plot all these  functions normalized to $1$  at $w=1$. We can see that $h_{+}$ (calculated with $m_{D^*}$), $h_{A_1}$, $h_{V}$ and $h_{A_3}$
are identical, even  when we would not  expect it from the different expressions  in  Eq. \eqref{eq:h}.
We can then see that our formalism implements exactly the symmetry of heavy quark physics, and provides an $w$ dependence for these functions.

 It is interesting to compare our results with those of  \cite{nieves}.  There a quark model calculation is done. and the quark matrix elements  are evaluated, including the transition
 form factor from $B$ to $D^*$  which we do not evaluate  with the claim that it cancels in ratios of amplitudes for different $M'$. We see that $h_{+}$ in \cite{nieves} is qualitatively similar
 to ours, although it falls faster with $w$. The difference with us are of the order of $15\%$ at the maximum value of $w$, indicating in any case a soft transition  matrix element.

 Next, in order to connect with the standard model we follow the formalism  of  \cite{caprini,Fajfer}
\begin{eqnarray}
&\frac{<D^*,\lambda, P_{D^*}|{\cal J}_\mu (0)|B, P_B>}{\sqrt{m_B m_{D^*}}} = \frac{2i V(q^2)}{m_B+m_{D^*}} \epsilon_{\mu\nu\alpha\beta} (\epsilon^{(\lambda)\nu})^* P^{\alpha}_B P^{\beta}_{D^*}
- 2 m_{D^*} A_0(q^2) \frac{\epsilon^{(\lambda)*} \cdot q}{q^2} q_\mu  \nonumber \, \\
& - ({m_B +m_{D^*}})  A_1(q^2) \left[\epsilon^{(\lambda)^*}_\mu -\frac{\epsilon^{(\lambda)^*} \cdot q}{q^2} q_\mu \right]    \nonumber \, \\
& + A_2(q^2) \frac{\epsilon^{(\lambda)*} \cdot q}{m_B+m_{D^*}}  \left[ (P_B+P_{D^*})_\mu -\frac{m^2_B-m^2_{D^*}}{q^2}q_\mu  \right]  \,,
\end{eqnarray}
where $q_\mu=P_{B\mu}-P_{D^*\mu}$.  Once again, comparing this expression with our results for $\mu=0$, $\mu=1,2,3$ with  $M'=0,+1,-1$, we obtain the following results:
\begin{eqnarray} \label{eq:s1}
V(q^2)&=&A A' (B-B') \frac{m_B +m_{D^*}}{2(E_{D^*}-E_B)}   \,,
\end{eqnarray}
\begin{eqnarray}\label{eq:s2}
A_0(q^2)&=&\frac{1}{2}A A' (B+B')   \,,
\end{eqnarray}
\begin{eqnarray}\label{eq:s3}
A_1(q^2)&=& \frac{1}{m_B +m_{D^*}} A A' (1-BB'p^2)     \,.
\end{eqnarray}

\begin{eqnarray}\label{eq:s4}
&(m_B +m_{D^*})A_1(q^2) \frac{E_{D^*}}{m_{D^*}} -\frac{2p^2(E_B-E_{D^*})}{m_{D^*}(m_B +m_{D^*})} A_2(q^2) \nonumber \, \\
&= A A' (1+B B' p^2) \,,~~
\end{eqnarray}
 from where we find
\begin{eqnarray}   \label{eq:s5}
A_2(q^2) &=&\frac{-m_{D^*}(m_B +m_{D^*})}{2(E_B-E_{D^*})}A A'  \nonumber \, \\
&\times& \left\{2 BB'-\frac{1}{p^2} \left(\frac{E_{D^*}}{m_{D^*}}-1\right) (1- BB'p^2) \right\}  \,.   ~~~~
\end{eqnarray}
As in  \cite{Fajfer} (Eq.(B.5)), we define here  $h_{A_1}(w)$  as
\begin{eqnarray}\label{eq:s6}
h_{A_1}(w)=\frac{2}{w+1} \frac{1}{R_{D^*}} A_1(q^2) \,,
\end{eqnarray}
with $R_{D^*}= \frac{2 \sqrt{m_B m_{D^*}}}{m_B+m_{D^*}}$. Hence  $h_{A_1}(w)$  here is identical to $h_{A_1}$  of  Eq. \eqref{eq:h}.

\begin{figure}[ht!]
\includegraphics[scale=0.72]{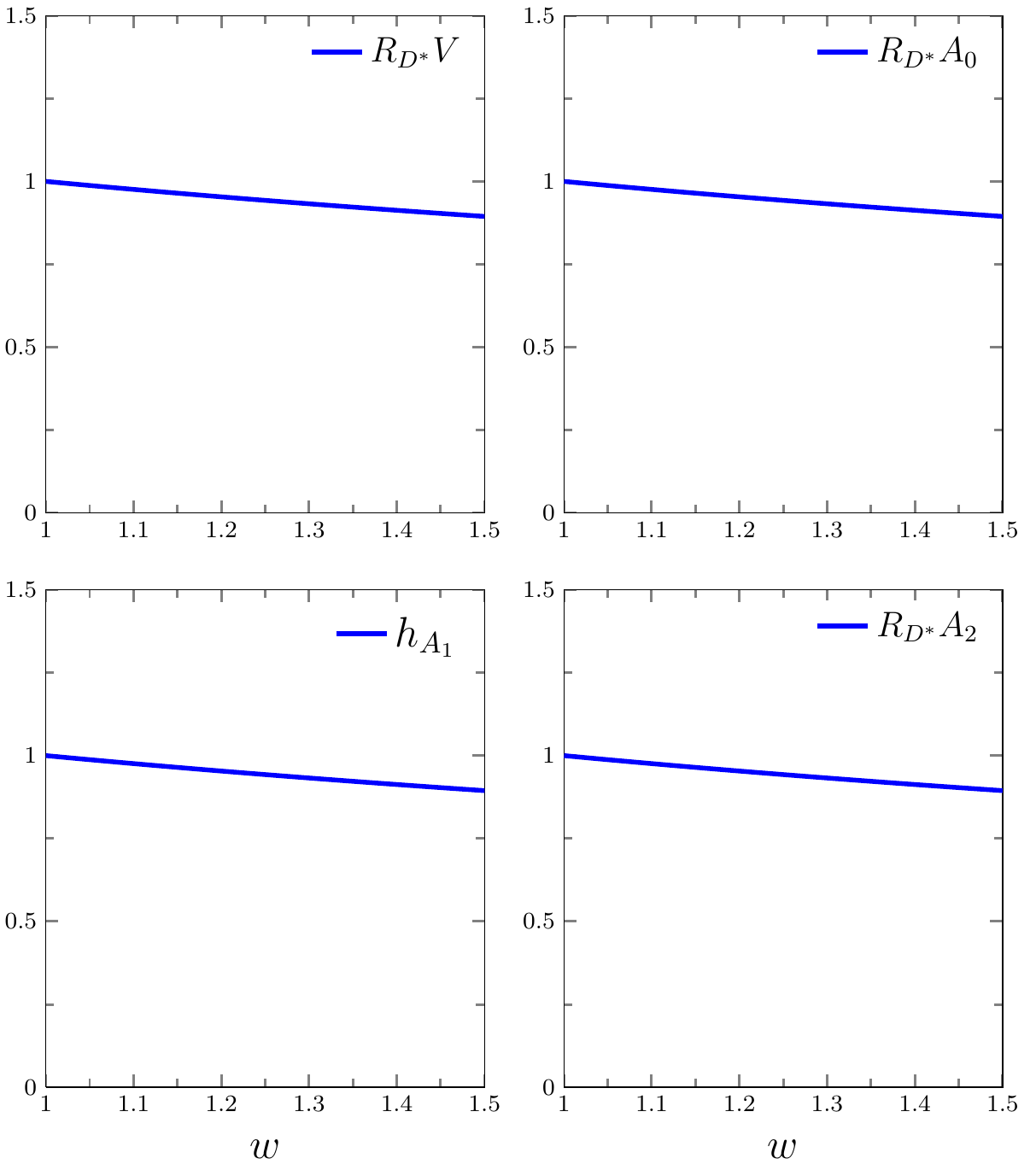}
\caption{$R_{D^*}V$, $R_{D^*} A_0$, $h_{A_1}$  and $R_{D^*} A_2$  from Eqs.  \eqref{eq:s1}, \eqref{eq:s2}, \eqref{eq:s5}, and \eqref{eq:s6}   as a function of $w$  normalized to $1$  at $w=1$. }
\label{fig:rv}
\end{figure}

In \cite{caprini,Fajfer} the $A_i, V$ form factors are parameterized as
\begin{eqnarray} \label{eq:f1}
A_0(q^2) =\frac{R_0(w)}{R_{D^*}} h_{A_1}(w) \nonumber \,,\\
A_2(q^2) =\frac{R_2(w)}{R_{D^*}} h_{A_1}(w) \nonumber \,, \\
V(q^2) =\frac{R_1(w)}{R_{D^*}} h_{A_1}(w) \,.
\end{eqnarray}
Our expressions in Eqs. \eqref{eq:s1},\eqref{eq:s2},\eqref{eq:s3},\eqref{eq:s4},\eqref{eq:s5} and \eqref{eq:s6} fulfill these conditions in the strict heavy quark  limit with
$R_0(w)=1$, $R_2(w)=1$, $R_1(w)=1$, such that $R_{D^*} A_i$ and $R_{D^*} V$  are exactly  equal to $h_{A_1}$. This is seen in Fig. \ref{fig:rv}. Diversions from the strict
heavy quark limit of the standard model  are incorporated in this formalism  parameterizing $h_{A_1}(w)$, $R_0(w)$, $R_1(w)$, $R_2(w)$  with the results  \cite{caprini,Fajfer}
\begin{eqnarray} \label{eq:f2}
&h_{A_1}(w)=h_{A_1}(1)[1-8\rho^2 z+(53\rho^2-15)z^2    \nonumber \, \\
&-(231\rho^2-91)z^3]\nonumber \,,  \\
&R_1(w)=R_1(1)-0.12(w-1)+0.05(w-1)^2 \nonumber \,,\\
&R_2(w)=R_2(1)+0.11(w-1)-0.06(w-1)^2 \nonumber \,,\\
&R_0(w)=R_0(1)-0.11(w-1)+0.01(w-1)^2  \,,
\end{eqnarray}
where  $z=\frac{\sqrt{w+1}-\sqrt{2}}{\sqrt{w+1}+\sqrt{2}}$,  with
\begin{eqnarray}\label{eq:f3}
R_0(1)=1.14   \nonumber \,,\\
R_1(1)=1.401 \pm 0.034 \pm 0.018  \nonumber \,,\\
R_2(1)=0.864 \pm 0.024 \pm 0.008 \nonumber \,,\\
\rho^2=1.214 \pm 0.034 \pm 0.009  \nonumber \,,\\
h_{A_1}(1)= 0.921 \pm 0.013 \pm 0.020 \,.
\end{eqnarray}
\begin{figure}[ht!]
\includegraphics[scale=0.72]{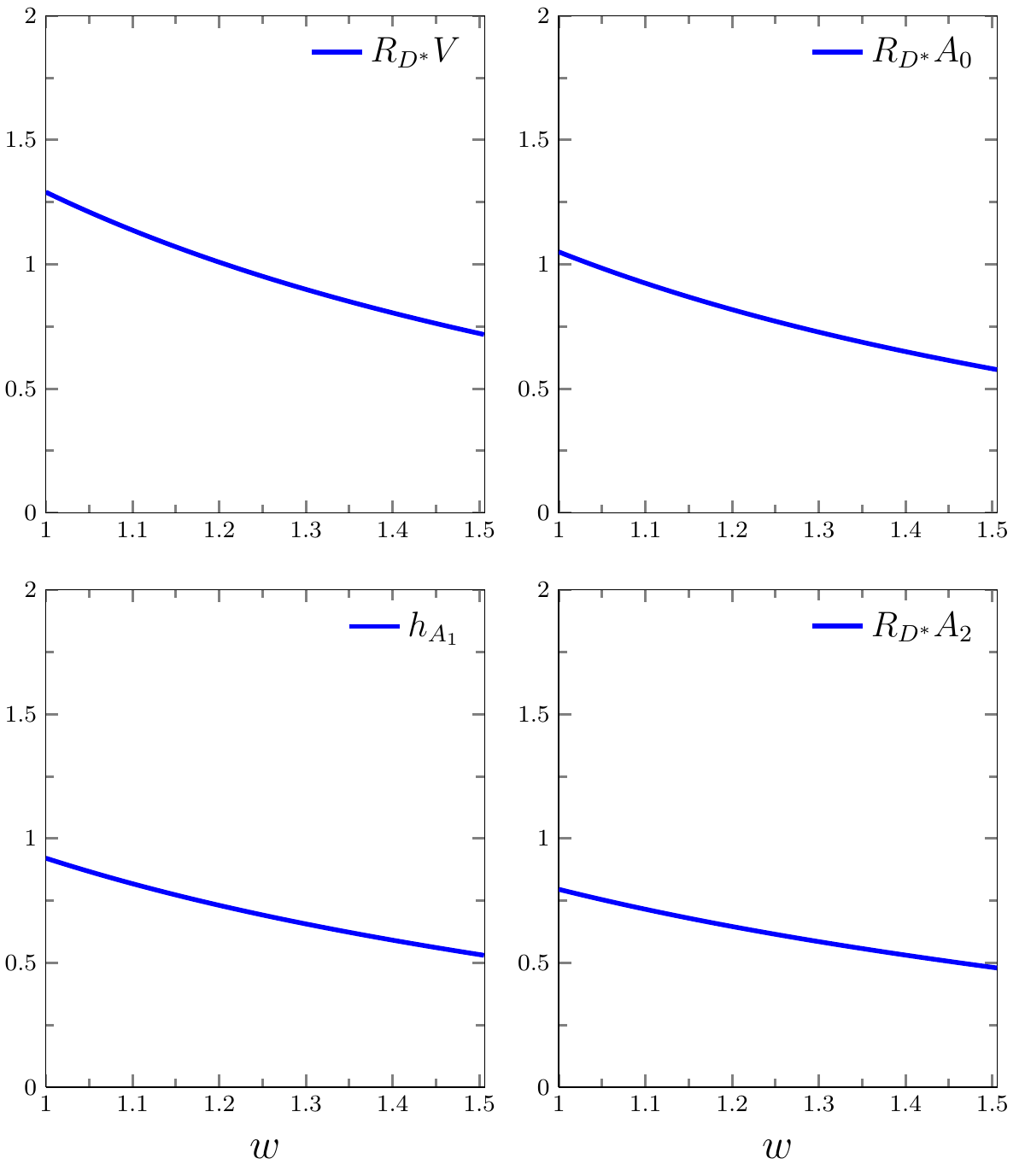}
\caption{the same as Fig. \ref{fig:rv2} but from Eqs.  \eqref{eq:f1}, \eqref{eq:f2} and \eqref{eq:f3}. }
\label{fig:rv2}
\end{figure}

\begin{figure}[ht!]
\includegraphics[scale=0.72]{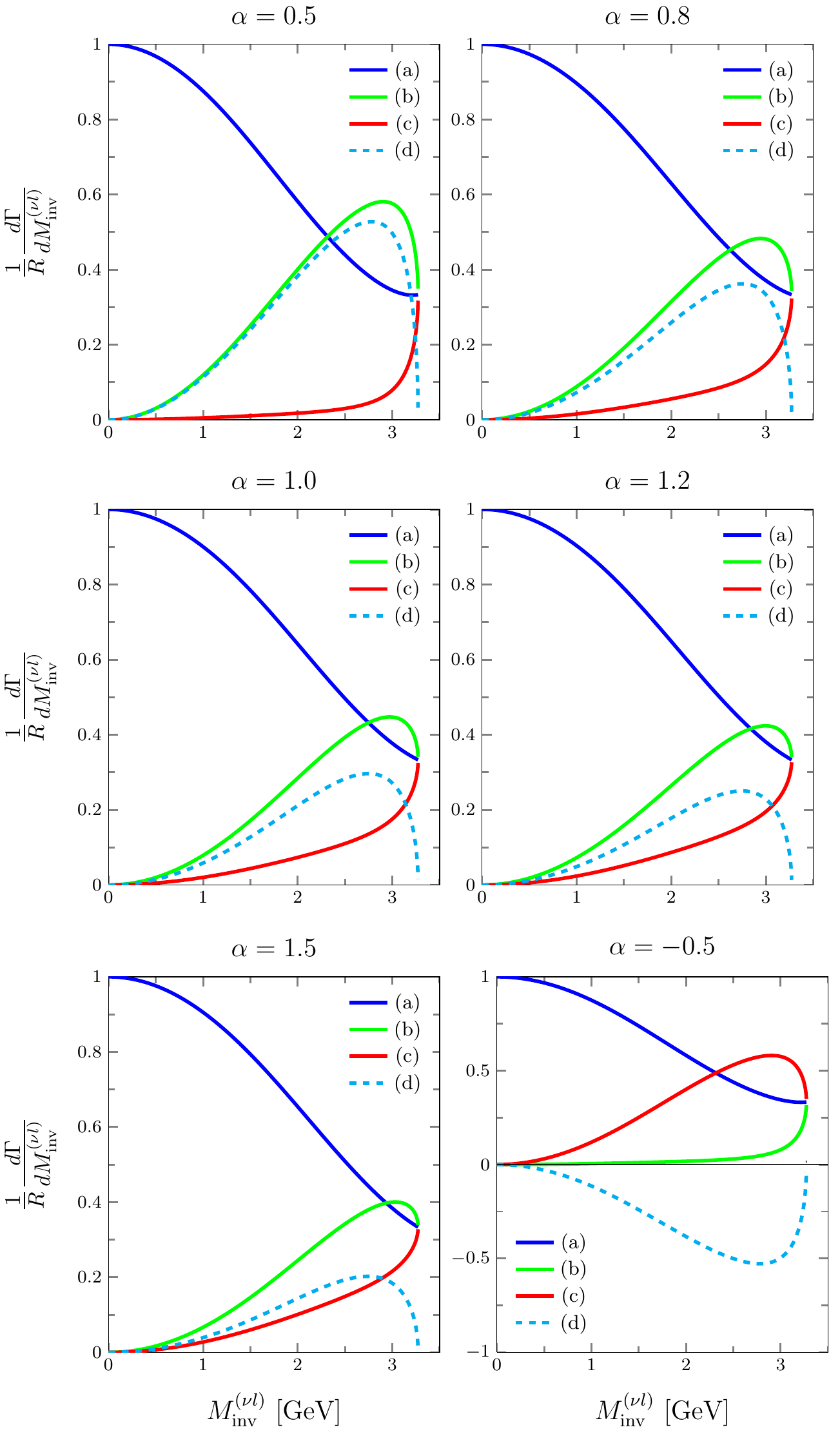}
\caption{The same as Fig. \ref{nfig:ro} but from Eqs.  \eqref{eq:f1}, \eqref{eq:f2}, \eqref{eq:f3},  and \eqref{eq:f4}. }
\label{fig:epjcdg}
\end{figure}

The results for  $h_{A_1}$ $R_{D^*}V$, $R_{D^*} A_0$, $R_{D^*} A_2$ are shown in Fig. \ref{fig:rv2}.  Comparison  of Fig. \ref{fig:rv} with Fig. \ref{fig:rv2} shows the  difference of our approach
 with the standard model. We can appreciate a bigger slope as a function of $w$ for the standard model (as already seen comparing with Ref. \cite{nieves}) and also a different normalization at $w=1$. Yet,
 the claim from our approach is that differences become much smaller  when we use our approach to calculate ratios of amplitudes. To see the accuracy of our model  to provide ratios, we  evaluate
 again the contribution of $M'=0,\pm 1$, divided by the sum of the three contributions, for different values of $\alpha$,  with the form factor of the standard model  and compare  the results
 with those obtained in Fig. \ref{nfig:ro}.  To evaluate those contributions in the standard model we look at the formulas of Eqs. \eqref{neq:tM0}, \eqref{neq:tM1},\eqref{neq:tMm1}, and looking at
 the expressions of Eqs. \eqref{eq:s1},\eqref{eq:s2},\eqref{eq:s3},\eqref{eq:s4},\eqref{eq:s5} and \eqref{eq:s6} we substitute,
\begin{eqnarray}\label{eq:f4}
& AA'(B+B')\,p   \to 2 A_0 p \nonumber \,,\\
& AA'(1+B B' p^2)  \to  \frac{E_{D^*}(m_B+m_{D^*})}{m_{D^*}} A_1 -\frac{2 p^2 (E_B-E_{D^*})}{m_{D^*}(m_B+m_{D^*})} A_2  \nonumber \,,\\
& AA'(1 - B B' p^2)   \to (m_B+m_{D^*}) A_1 \nonumber \,,\\
& AA'(B -B')\,p \to \frac{2(E_{D^*}-E_B)}{(m_B+m_{D^*})} V p  \,.
 \end{eqnarray}
The results are shown in Fig. \ref{fig:epjcdg}. One can appreciate some differences from Fig. \ref{nfig:ro}, but they are very small. For $M_{\rm inv}^{(\nu l)}$ maximum,  which corresponds to $w=1$
the results are practically identical. The differences are more  visible for small $M_{\rm inv}^{(\nu l)}$, a region which  is anyway suppressed  by phase space in the mass distributions.
The fact that the three contributions are equal at $w=1 (M_{\rm  inv}|_{\rm max})$  in both approaches is trivial since only $A_1$  contributes there and the expressions for $\sum |t|^2$ in terms of $A_1$ are identical
for all $M'$. The fact that close to $M_{\rm  inv}|_{\rm max}$  the behaviour  in both cases is so close can also be traced to the fact that for  a certain  range  of $p$ momentum the $A_1$ term
is still largely dominant. Yet, this could be seen as a manifestation  of a general behaviour of the helicity amplitudes close to the end point discussed in Ref. \cite{hiller}.

\section{Conclusions}
We have taken advantage of a recent reformulation of the weak decay of hadrons, where, instead of parameterizing the amplitudes in terms of particular structures  with
their corresponding  form factors, the weak transition matrix elements at the quark level  are mapped into hadronic matrix elements and an elaborate angular momentum algebra is
performed  that allows one to correlate  the decay amplitudes for a wide range of reactions. The formalism  allows  one to  obtain easy  analytical formulas for each reaction  in
terms of the angular momentum components of the hadrons.  One global  form factor  also appears  in the approach related  to the radial  wave functions of the hadrons involved, but since
this  form factor  is common to many reactions and  in particular is exactly  the same for the different  spin components of the hadrons within the same reaction, it cancels in ratios of amplitudes
or differential mass distributions.

 In the present paper we have taken  this formalism and extended  it to the case  of hadron matrix elements with an operator $\gamma^\mu -\alpha\gamma^\mu \gamma_5$, which can
 accommodate many  models beyond the standard model by changing $\alpha$.  We have applied the formalism  to study the
 $B \to  D^{*} \bar{\nu} l$ reaction  and the amplitudes  for different  helicities of the $D^*$ are evaluated. We see that   $\frac{d \Gamma}{d M_{\rm inv}^{(\nu l)}}$  depends
 strongly  on  the helicity amplitude and also on the $\alpha$ parameter. In particular the difference   $\frac{d \Gamma}{d M_{\rm inv}^{(\nu l)}}|_{M=-1} -\frac{d \Gamma}{d M_{\rm inv}^{(\nu l)}}|_{M=+1} $
 is shown to be very sensitive to the  $\alpha$ parameter and changes  sign when  we go from  $\alpha$ to $-\alpha$. Such a magnitude, with its strong sensitivity to this  parameter, should
 be an ideal test to investigate  models beyond the standard model and we encourage its measurement in this and analogous reactions, as well as the theoretical calculations for different models.

 We have taken advantage to relate our approach to the standard model by calculating  the  form factors $V(q^2)$, $A_0(q^2)$, $A_1(q^2)$, $A_2(q^2)$ in our approach  and comparing them to the
parameterization  of the  standard model.  The form factors are qualitatively  similar but one can observe differences. Yet,  when one uses them to evaluate ratios of amplitudes, or partial differential
mass distributions,  the differences are very small, and near the end point $w=1$ the distributions are practically  identical.

\section*{Acknowledgments}
We wish to express our thanks to Jose Valle, Martin Hirsch, Avelino Vicente and Xiao-Gang He for useful discussions.
Also  discussions with Juan Nieves and Eliecer Hernandez are much appreciated.
LRD acknowledges the support from the National Natural Science
Foundation of China (Grant No. 11575076) and the State Scholarship Fund of China (No. 201708210057).
This work is partly supported by the Spanish Ministerio
de Economia y Competitividad and European FEDER funds under Contracts No. FIS2017-84038-C2-1-P B
and No. FIS2017-84038-C2-2-P B, and the Generalitat Valenciana in the program Prometeo II-2014/068, and
the project Severo Ochoa of IFIC, SEV-2014-0398 (EO).



\end{document}